\documentclass[10pt,a4]{article}
\usepackage{amsmath,amssymb}

\usepackage{graphicx}
\usepackage{natbib}
\usepackage[hyphens]{url}

\newcommand{\real}{\mathbb{R}}

\date{}
\begin{document}

\title{A new class of spatial covariance functions generated by higher-order kernels}
\maketitle
  
\begin{center}
  {\small\bf Mohammad Ghorbani\footnote{Corresponding Author} }\\\thanks{Department of mathematics and mathematical statistics, Ume\r{a} University, Sweden\\E-mail: mohammad.ghorbani@umu.se} 

  {\small\bf Jorge Mateu}\\
  \thanks{Department of Mathematics, Jaume I University, Castell\'{o}n, Spain\\E-mail: mateu@uji.es}\\   

\end{center}
\baselineskip=0.5cm
\begin{abstract}
Covariance functions and variograms  play a fundamental role for exploratory analysis and statistical modelling of spatial and spatio-temporal datasets. In this paper, we construct a new class of spatial covariance functions using the Fourier transform of some  higher-order kernels.
Further, we extend  this class of the spatial covariance functions to the spatio-temporal setting by using the idea used in \cite{ma:03}.

\end{abstract}

\section{Introduction}
Suppose that  $\{Z(u); u\in D\subset \mathbb{R}^d\}$ is a spatial random process observed at $n$ locations $u_1,\ldots,u_n$. In most practical applications  $d=2$ or $3$. 
In analyzing spatial data, usually the main objective is the optimal prediction of  unobserved parts of the process on the basis of the  observations. To achieve this goal, we require a suitable model to reveal  how observations co-vary with respect to each other in space. In other words, a suitable covariance or variogram model is needed. Thus, a  fundamental concept in drawing inferences from  data related to the phenomena under study is a covariance or a variogram function, and a new class of models is a welcome contribution in analyzing spatial data.

We now review a few important concepts of spatial statistics.
Throughout  the paper, we assume that the spatial process $Z(\cdot)$ satisfies the regularity condition, $\mathbb{V}ar[Z(u)]<\infty$, for all $u\in D$, which implies that the first two moments exist. By this assumption, at each location point $s$ in $D$ we can define the mean  function as
\[\mu(u)=\mathbb{E}[Z(u)]\]
 and for every two location points $u_1$ and $u_2$ in $D$, the  covariance and the semivariogram functions can respectively be defined as \[C(u_1, u_2)=Cov[Z(u_1), Z(u_2)]\]
 and \[\gamma(u_1, u_2)=\frac{1}{2}{\mathbb{V}}ar[Z(u_1)-Z(u_2],\]
 provided that they exist.
We also assume that the process $\{Z(u)\}$ is second-order stationary, meaning that the mean function is constant and the covariance function depends on the difference between two distinct points only, i.e., for some function $C_0$, \[C(u_1, u_2)=C_0(u_1-u_2).\]

The function $C_0$ is a valid covariance function if it is even and satisfies the positive definiteness condition. That is, for any $u_1,\ldots,u_m$ and reals $a_1, a_2, \ldots, a_m$, and any positive integer $m$, $C_0$ must satisfy 
 \[\sum_{i=1}^m\sum_{j=1}^m a_ia_jC_0(u_i-u_j)\ge 0.\]
For a continuous  covariance function $C_0$ evaluated at spatial lag  $ h=u_i-u_j$ this is equivalent to the  Bochner's theorem, stating that $C_0$ is positive definite if and only if it can be represented as
\[C_0(h)=\int e^{ih^{\mathsf{T}}\omega}\mathrm{d}F(\omega),\]
where $F$ is a non-decreasing, right continuous, and bounded real-valued function which is called the spectral measure of $C_0(h)$   \citep{Ripley:81,Lindgren:12} and $h^{\mathsf{T}}$ is the transpose of $h$. If
$F$ is absolutely continuous with respect to the Lebesgue measure, then $\mathrm{d}F(\omega) = f(\omega)\mathrm{d}\omega$ and $f(\omega)$ is called the spectral density.
It means that, for any absolutely continuous function $F$, the covariance function $C_0$ is its characteristic function.

There is another type of stationarity, called intrinsic stationarity, which is based on
the variogram, and it is more general than second-order stationarity since there are processes
for which the variogram is well defined but the covariance is not. The process $Z(s)$ is said to be intrinsically stationary, if
its mean function is constant
and the variogram depends only on the spatial distances $u_1-u_2$ for every $u_1, u_2 \in \mathbb{R}^d$.
The corresponding semivariogram for some function $\gamma_0$  is denoted by $\gamma_0(u)$ and said to be an intrinsically stationary semivariogram. If the process $Z(u)$ is second-order stationary, then it is also intrinsically stationary
and
\begin{align}
\label{eq:covSem}
\gamma(u) = C(0) - C(u),\qquad u\in\real^d.
\end{align}

Further, a second-order  stationary random process $Z(\cdot)$ is said isotropic if its covariance function $C(h)$ (or the variogram function $\gamma(h)$) only depends on $\|h\|$, where $\|.\|$ indicates the Euclidean distance. Hereafter, we assume that  the process $Z(\cdot)$ is isotropic.
To assure positive definiteness, it is usually assumed that the  covariance function $C_0$  belongs to a parametric family whose members are known to be positive definite. That is, one assumes that 
\begin{align}
\label{eq:1}
C_0(h)=C^0( h;\theta),
\end{align}
where $C^0$ satisfies the positive definiteness condition for all $\theta\in\Theta\subset \mathbb{R}^d$. The vector parameter $\theta$ usually consists of one or more of the following parameters: the nugget effect, the sill,  the partial sill, the  range, and the smoothing parameter.
 There are several commonly used parametric models such as  exponential, Gaussian, spherical, power exponential, Cauchy, and Mat\'{e}rn in the literature for the variogram and covariance functions in geostatistical modelling, 
see, e.g. \cite{cressie:93,Ripley:81}.
In this paper, we introduce  new  classes  of  spatial and spatio-temporal covariance functions by means of the characteristic functions of absolutely continuous higher-order kernels. The plan of the paper is the following. Section~\ref{sec:SpCov} presents new covariance functions for spatial dimension, while Section~\ref{sec:STCov} does the same for the spatio-temporal case. Section~\ref{sec:rainfall} presents the analysis of Swiss rainfall data. The paper ends with some  conclusion.
\section{Spatial covariance functions generated by higher-order kernels}\label{sec:SpCov}
We start this section by introducing some notation.
A function $K(x)$ is called a symmetric kernel function if $K(x)=K(-x)$ and $\int_{\mathbb{R}} K(x)\mathrm{d}x=1$. Further, the kernel function $K$ is called s-smooth if for $s\ge 1$, its $(s-1)$th derivative, i.e. $K^{(s-1)}$, is absolutely continuous on $\mathbb{R}$. For any integer $j\ge 1$, consider the notation  $\mu_j(K) =\int_{\mathbb{R}}x^jK(x)\mathrm{d}x$. We call a kernel function $K$ is of order $2r$ if $ \mu_j(K)=0$ for all $j<2r$ and $\mu_{2r}(K)\ne 0$.

A special function which we will use throughout the paper is the spherical Bessel function. For any integer $m\ge 0$, it is given by 
\begin{align}
\label{eq:2}
j_m(x)=\sum_{k=0}^\infty\frac{(-1)^k}{\left(\frac{3}{2}\right)_{m+k}k!}\left(\frac{x}{2}\right)^{m+2k}.
\end{align}
Here $\left(\frac{3}{2}\right)_{m+k}$ obeys Pochhammer's symbol  given by
\[(\alpha)_n=\alpha(\alpha+1)(\alpha+2)\ldots(\alpha+n-1)=\frac{\Gamma(\alpha+n)}{\Gamma(\alpha)},\] 
where $\Gamma(\cdot)$ is the gamma function.
\subsection{M\"{u}ller's kernel}
We now consider a new class of $s$-smooth bounded-support kernels of order $2r$ introduced by \cite{muller:84}.  This class of s-smooth, $2r$th-order kernel on [-1,1]  minimizes the mean integrated square error in kernel density estimation. \cite{hansen:05} presented an alternative representation of M\"{u}ller's kernel functions. We use Hansen's representation of an s-smooth, $2r$th-order kernel for building  a new class of spatial covariance functions.
Following \cite{hansen:05}, for any integer $r\ge 1$, the M\"{u}ller's s-smooth, $2r$th-order kernel on [-1,1]
is given by
\begin{align}
\label{eq:3}
M_{2r,s}(x)=B_{r,s}(x)M_s(x),
\end{align}
where
\begin{align*}
B_{r,s}(x)=\frac{\left(\frac{3}{2}\right)_{r-1}\left(\frac{3}{2}+s\right)_{r-1}}{(s+1)_{r-1}}\sum_{k=0}^{r-1}\frac{(-1)^k\left(\frac{1}{2}+s+r\right)_{k}x^{2k}}{k!(r-1-k)!\left(\frac{3}{2}\right)_{k}}, 
\end{align*}
and for $s\ge 0$
\begin{align*}
M_s(x)=\frac{\left(\frac{1}{2}\right)_{s+1}}{s!}(1-x^2)^s.
\end{align*}
Note that $M_s(x)$ is a special case of $M_{2r,s}(x)$ for the case $r=1$. \cite{granovsky:muller:91} showed that 
\begin{align}
\label{eq:4}
\lim_{s\rightarrow\infty}\frac{1}{\sqrt{2s}} M_{2r,s}\left(\frac{x}{\sqrt{2s}}\right)=G_{2r}(x)=\frac{(-1)^r\phi^{(2r-1)}(x)}{2^{r-1}(r-1)!x},
\end{align}
where $\phi^{(2r-1)}(x)$ is the $(2r-1)$th derivative of a standard normal density, and $G_{2r}$ is the higher-order Gaussian kernel.
\subsection{Spatial covariance functions}
We construct here a family of spatial covariance functions by using the characteristic function of the above kernels. Toward this end, for any function $g$, we denote its characteristic function by $\tilde{g}$. 
\subsubsection{Higher-order Gaussian covariance functions}
\begin{figure}[!htbp]
\begin{center}
\includegraphics[width=11cm,height=6cm]{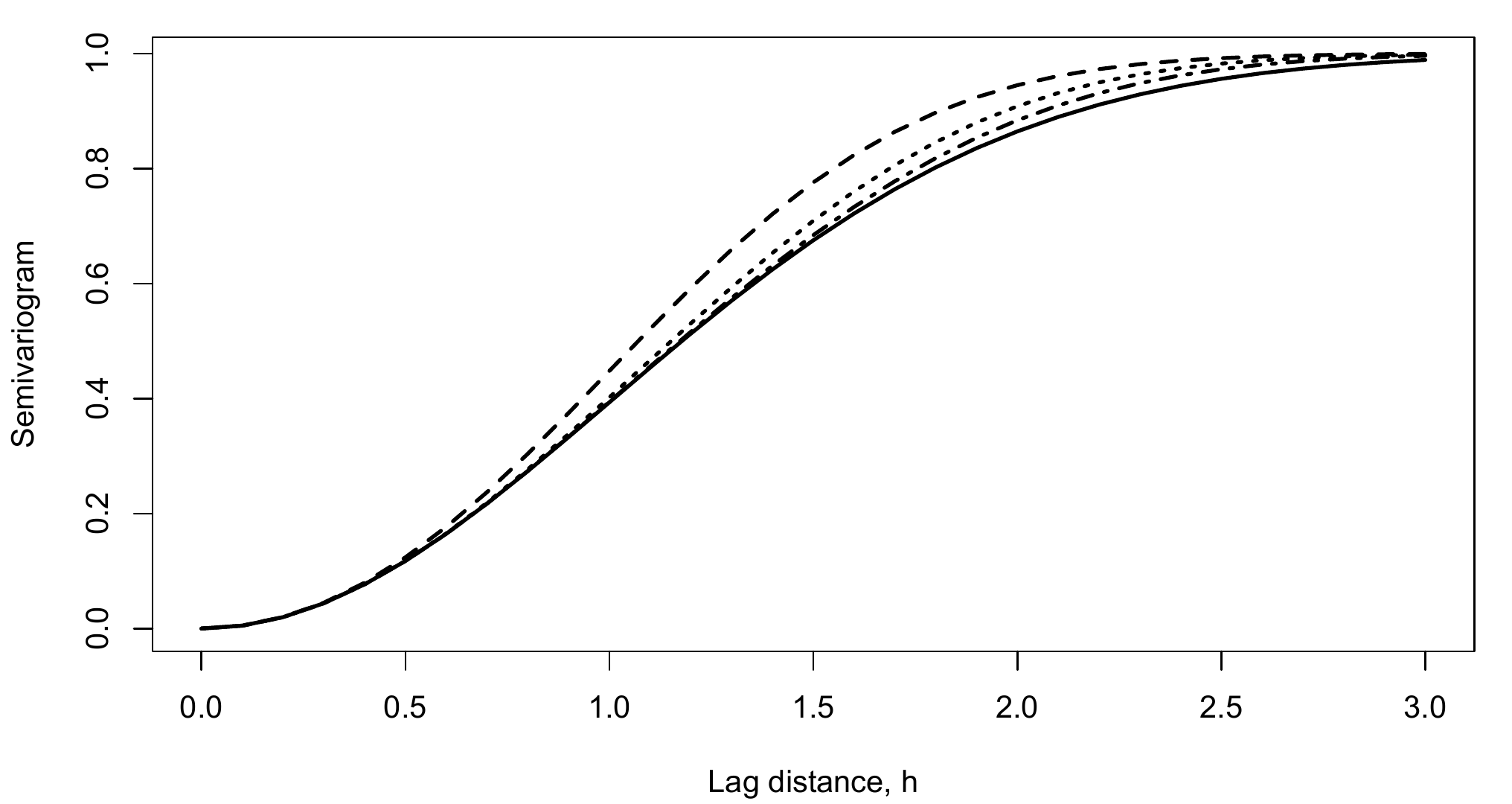}
\caption{Isotropic stationary higher-order Gaussian covariance functions: Gaussian (solid line), Gaussian with order=2 (dashed line), Gaussian with order=3 (dotted line), and Gaussian with order=4 (dot-dashed line).}
\label{fig:1}
\end{center}
\end{figure}

For the higher-order Gaussian kernels given in \eqref{eq:4}, the characteristic function is given by $\tilde{G}_{2r}(h)=\exp(-h^2/2)\sum_{k=0}^{r-1}h^{2k}/(2^kk!)$, which introduce a new class of spatial covariance functions. Particularly, for $r=1$, $\tilde{G}_{2r}(h)= \exp(-h^2/2)$ is the classical Gaussian covariance model which we denote by $\Phi(h)$, see, e.g.\ \cite{wackernagel:95}. Isotropic stationary higher-order Gaussian covariance functions for some special values of $r$ are listed in Table~\ref{t:1}.
The set of functions in Table~\ref{t:1} are represented in Figure~\ref{fig:1}. They have the same behavior on two tails but they are slightly different along the support.

\begin{table}
\centering{
\caption{Isotropic stationary higher-order Gaussian covariance functions}
\begin{tabular}{ccccc}
\hline
r &1 &2 &3 &4 \\
\hline
$\tilde{G}_{2r}(h)$& $\Phi(h)$&$\Phi(h)(1+h^2/2)$ &$\Phi(h)(1+h^2/2+h^4/8)$ &$\Phi(h)(1+h^2/2+h^4/8+h^6/48)$\\
\hline
\end{tabular}

\label{t:1}
}
\end{table}

We note that the family of higher-order Gaussian covarince functions can be generalized by considering {\it Laguerre polynomials}, see e.g.\, \cite{fasshauer:07}. We leave such extensions to be
discussed in future work.
\subsubsection{M\"{u}ller's covariance functions}
Another class of stationary spatial covariance function is obtained by using the characteristic function of M\"{u}ller's s-smooth, $2r$th-order kernel. The characteristic function of the M\"{u}ller's higher-order kernel is given by
\begin{align}
C_1(h)=\tilde{M}_s(h)=\left(\frac{3}{2}\right)_s\left(\frac{2}{h}\right)^sj_s(h)
\label{eq:4'}
\end{align}
and 
\begin{align}
\label{eq:5}
C_2(h)=\tilde{M}_{2r,s}(h)=\frac{2}{\sqrt{\pi}}\left(\frac{2}{h}\right)^s\sum_{m=0}^{r-1}\alpha_s(m)j_{s+2m}(h),
\end{align}
where
\begin{align*}
\alpha_s(m)=\frac{\Gamma\left(\frac{1}{2}+m+s\right)\left(\frac{1}{2}+2m+s\right)}{m!}.
\end{align*}
See more details in \cite{hansen:05}. 
It is easy to show that \eqref{eq:4'} is another representation of the Bessel covariance function  \cite[p.~139]{yaglom:87}.  
Further, it can be shown that the Mat\'{e}rn covariance function \citep{matern:60} which is related to the characteristics function of the student's t distribution \citep{hurst:95,ghorbani:17} is  a special case of \eqref{eq:5}.
Thus \eqref{eq:5} provides a general class of covariance functions for different values of $r$.

For computational purpose, we use the recurrence formula of the Bessel function, $j_m(h)$, given by
\begin{align*}
j_{m+1}(h)=\frac{2m+1}{h}j_m(h)+j_{m-1}(h)
\end{align*}
with the initial condition
\begin{align*}
j_0(h)=\frac{\sin(h)}{h}
\end{align*}
 and
 \begin{align*}
 j_1(h)=\frac{\sin(h)}{h^2}-\frac{\cos(h)}{h}.
 \end{align*}
In the special case by setting $r=1$ and $s=0$ in \eqref{eq:5}, the  hole effect
(in some literature called the sine wave) model, 
\begin{align*}
\tilde{M}_{2,0}(h)= \frac{\sin(h)}{h},
\end{align*}
 is obtained. Note that the hole effect model is the characteristic function of a continuous uniform random variable on $[-1,1]$. This model can be reparametrized as
 \begin{align*}
 C({\bf h}, \theta)=\sigma_e^2+\sigma^2\frac{\eta}{\|h\|}\sin\left(\|h\|/{\eta}\right),\qquad {\bf h \ne 0}, 
 \end{align*}
$\theta= (\sigma_e^2, \sigma^2, \eta)$, where $\sigma_e^2\ge 0$ , $\sigma^2\ge 0$ and $\eta\ge 0$ denote the nugget effect, the sill and the smoothing parameter, respectively.
Considering \eqref{eq:covSem}, the corresponding semivariogram is
given by 
 \begin{align*}
 C({\bf h}, \theta)=\sigma_e^2+\sigma^2\left(1-\frac{\eta}{\|h\|}\sin\left(\|h\|/{\eta}\right)\right),\qquad {\bf h \ne 0}, 
 \end{align*}

  This model is used for modelling the data where the empirical variogram shows strong cyclicity with decreasing amplitudes for increasing lag distances. In the literature, so far, this was the only example of a process where the covariance has a ciclicity behavior as a function of the distance $h$,
and can be directly obtained from the characteristic function of a kernel function. 
Using  this class of covariance models we are able to generate other processes where the covariance function has a sinusoidal behavior but not as strong as a sine wave. For example, consider the model  $\tilde{M}_{2,1}(h)$ given by
 \[\tilde{M}_{2,1}(h)=\frac{3}{h}\left[\frac{\sin(h)}{h^2}-\frac{cos(h)}{h}\right].\] 
 A parametric version of this model is given by 
 \begin{align}
 \label{eq:sincos}
 \tilde{M}_{2,1}({\bf h}, \theta)=\sigma_e^2+\sigma^2\frac{3\eta}{\|h\|}\left[\sin\left(\frac{\|h\|}{\eta}\right)\frac{\eta^2}{\|h\|^2}-cos\left(\frac{\|h\|}{\eta}\right)\frac{\eta}{\|h\|}\right],
 \end{align}
 where $\sigma_e^2\ge 0,\, \sigma^2\ge 0,$ and $\eta\ge 0$.
\begin{figure}[!htbp]
\begin{center}
\includegraphics[width=12cm,height=7cm]{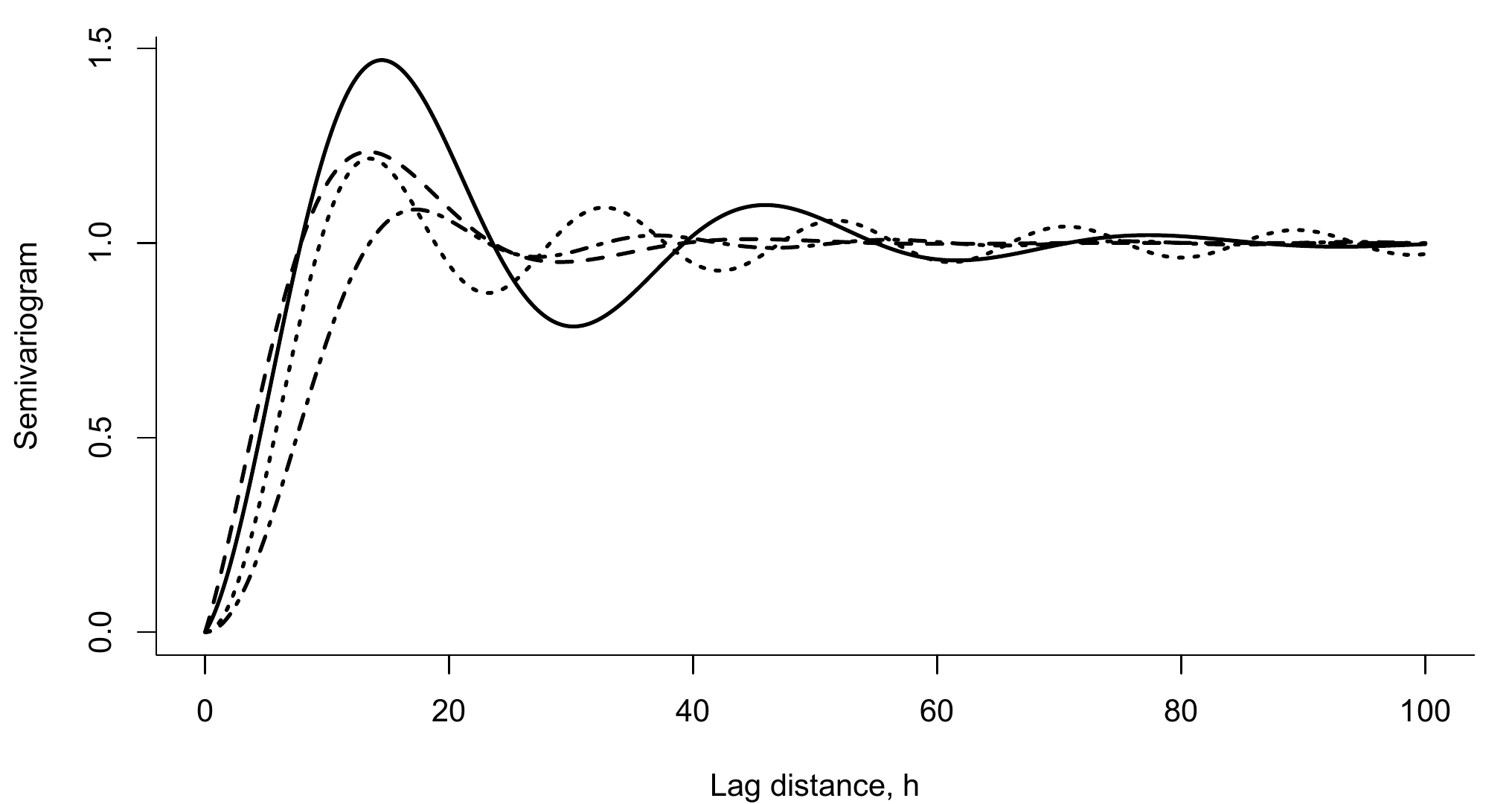}
\caption{Examples of isotropic stationary semivariogram functions: Exponential-cosine composite semivariogram with $\nu=60$ and $\eta=5$ (solid line) and   with $\nu=30$ and $\eta=5$ (dashed line). Sine-cosine wave with $\eta=3$ (dot-dashed line) and sine wave with $\eta=3$ (doted line). In all models $\sigma_e^2=0$ and $\sigma^2=1$.}
\label{fig:2}
\end{center}
\end{figure}
We call this model the {\it sine-cosine wave} model. As shown in Figure~\ref{fig:2} this model can be used in the situation where the cyclicity  weakens  because of great variation along the support.
Further, as it is clear Figure~\ref{fig:2}, one can use it in place of or along with the cosine-based composite covariance functions which are obtained by the product of the cosine covariance function and other positive definite covariance functions.  For instance, the {\it cosine-exponential} composite model
\begin{align*}
C({\bf h},\theta)=\sigma_e^2+\sigma^2\exp(-3\|h\|/\nu)cos(\|h\|/\eta),\qquad \sigma^2>0,\, \nu>0,\, \eta>0, 
\end{align*}
 is obtained by the product of the {\it cosine} and {\it exponential} covariance models, see more details in \cite{yaglom:87}, page 122.
\section{Spatio-temporal covariance functions}\label{sec:STCov}
Here we use the idea in \cite{cressie:huang:99} and \cite{ma:03} to extend our spatial covariance function to the case of spatio-temporal covariance function.
To this end, consider a real-valued random process $Z(u;t)$, indexed in space by $u\in \mathbb{R}^d$ and in
time by $t\in T$. As in the spatial case the spatio-temporal dependence is usually characterized by the covariance function
\begin{align}
C(u_1, u_2; t_1, t_2)=Cov[Z(u_1;t_1), Z(u_2; t_2)]
\label{eq:6}
\end{align}

 For $C(u_1, u_2; t_1, t_2)$ to be well-defined, we need to assume that $\mathbb{V}ar[Z(u;t)]<\infty$. The spatio-temporal stochastic process $Z(u;t)$ is called second-ordery stationary if the mean function $\mu(u;t)=\mathbb{E}[Z(u;t)]$ is constant and the covariance function \eqref{eq:6} is a function of the spatial distance $h=u_1-u_2$ and temporal lag $t=t_1-t_2$. Further, the process is called isotropic if $h=\|h\|$ and $t=t$. In this case, for some function $c^0$, we denote the covarince function by $C^0(h;t)$.
 
\begin{figure}[!htbp]
\begin{center}
\includegraphics[width=15cm,height=8cm]{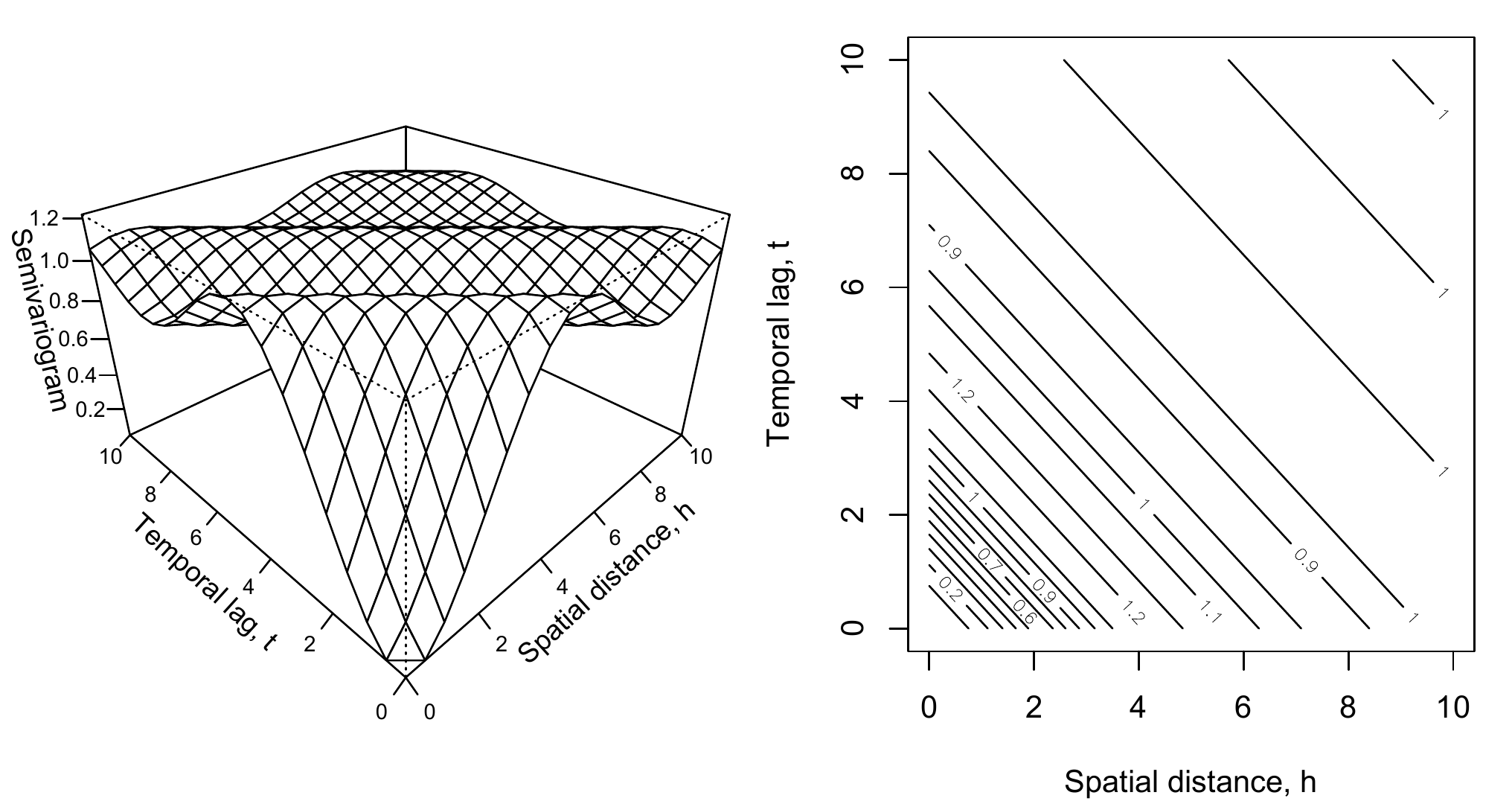}
\caption{Isotropic stationary spatio-temporal covariance function given in \eqref{eq:8} with $\beta=1$.}
\label{fig:3}
\end{center}
\end{figure}

 Corollary 1.3 in \cite{ma:03} states that for a constant  vector $\beta\in\mathbb{R}^d$, if $C_S(h)$ is a stationary covariance function on $\mathbb{R}^d$, then 
 \begin{align}
 C(h;t)=C_S(h+\beta t), \qquad (h;t)\in\mathbb{R}^d\times\mathbb{R}
\label{eq:7}
 \end{align}
is a stationary covariance on $\mathbb{R}^d\times\mathbb{R}$, provided that the expectation exists. The same idea with slightly different notation has been used in \cite{cressie:huang:99}.
Thus, using the above corollary, 
\begin{align*}
C_1(h;t)=C_1(h+\beta t)=\left(\frac{3}{2}\right)_s\left(\frac{2}{h+\beta t}\right)^sj_s(h+\beta t).
\end{align*}
and 
\begin{align*}
C_2(h;t)=C_2(h+\beta t)=\frac{2}{\sqrt{\pi}}\left(\frac{2}{h+\beta t}\right)^s\sum_{m=0}^{r-1}\alpha_s(m)j_{s+2m}(h+\beta t),
\end{align*}
are stationary spatio-temporal covariance functions.
In the spatial case 
\begin{align}
\label{eq:8}
\tilde{M}_{2,0}(h+\beta t)= \frac{\sin(h+\beta t)}{h+\beta t},
\end{align}
and
 \[\tilde{M}_{2,1}(h+\beta t)=\frac{3}{h+\beta t}\left[\frac{\sin(h+\beta t)}{h^2}-\frac{cos(h+\beta t)}{h+\beta t}\right]\] 
  are wave spatio-temporal covarince functions. Perspective and contour plots of \eqref{eq:8} are shown in Figure~\ref{fig:3}.
\section{Swiss rainfall data}\label{sec:rainfall}
In this section, we illustrate how our sine-cosine wave model applies to the Swiss rainfall data. The data are the record of rainfall measured on 8th May 1986 at 467 locations across Switzerland. 
 The data analyzed in this section is taken from 
\url{http://www.leg.ufpr.br/doku.php/pessoais:paulojus:mbgbook:datasets}.
This data collection was part of a workshop
organized by AI-GEOSTATS to compare the various contemporary methods in use of analyzing spatial data, see
\cite{dubois:98} for a detailed description of the data and project. 

We use the weighted least square (WLS) method {\sf explain why wee use this WLS method}
\begin{align*}
Q(\theta)=\sum_{i=1}^k\left[\log(2\hat{\gamma}(h_i))-\log(2\gamma(h_i,\theta))\right]^2 N(h_i)/2,
\end{align*}
 for parameter estimation, where $\hat{\gamma}(h)$ is the empirical semivariogram (see \cite{cressie:93}, page 69), $k$ denotes the number of lag distances at which the empirical and theoretical
semivariograms are computed, and $N(h)$ is the number of distinct pairs with distance $h$. We obtained the empirical semivariogram, $\hat{\gamma}(h)$, by using the R package {\bf geoR} \citep{ribeiro:diggle:01}.
To determine the WLS estimate of $\theta$,  the  R package {\bf nloptr} \citep{ypma:17} was used. First, we used the DIRECT-L method   to obtain a global optimum $\bar \theta$. Afterwards, to polish the optimum to a greater accuracy, we used $\bar\theta$ as a starting
point for the local optimization ‘bound-constrained by quadratic approximation’
(BOBYQA) algorithm \citep{powell:09} and obtained a final estimate $\hat\theta$.
The WLS estimates of the sine-cosine wave model~\eqref{eq:sincos} parameters 
when $\sigma_e^2=2766$ (by the empirical variogram) are $\hat{\sigma}=103.17$ and $\hat{\eta}=14.13$. 
\begin{figure}[!htbp]
\begin{center}
\includegraphics[width=12cm,height=7cm]{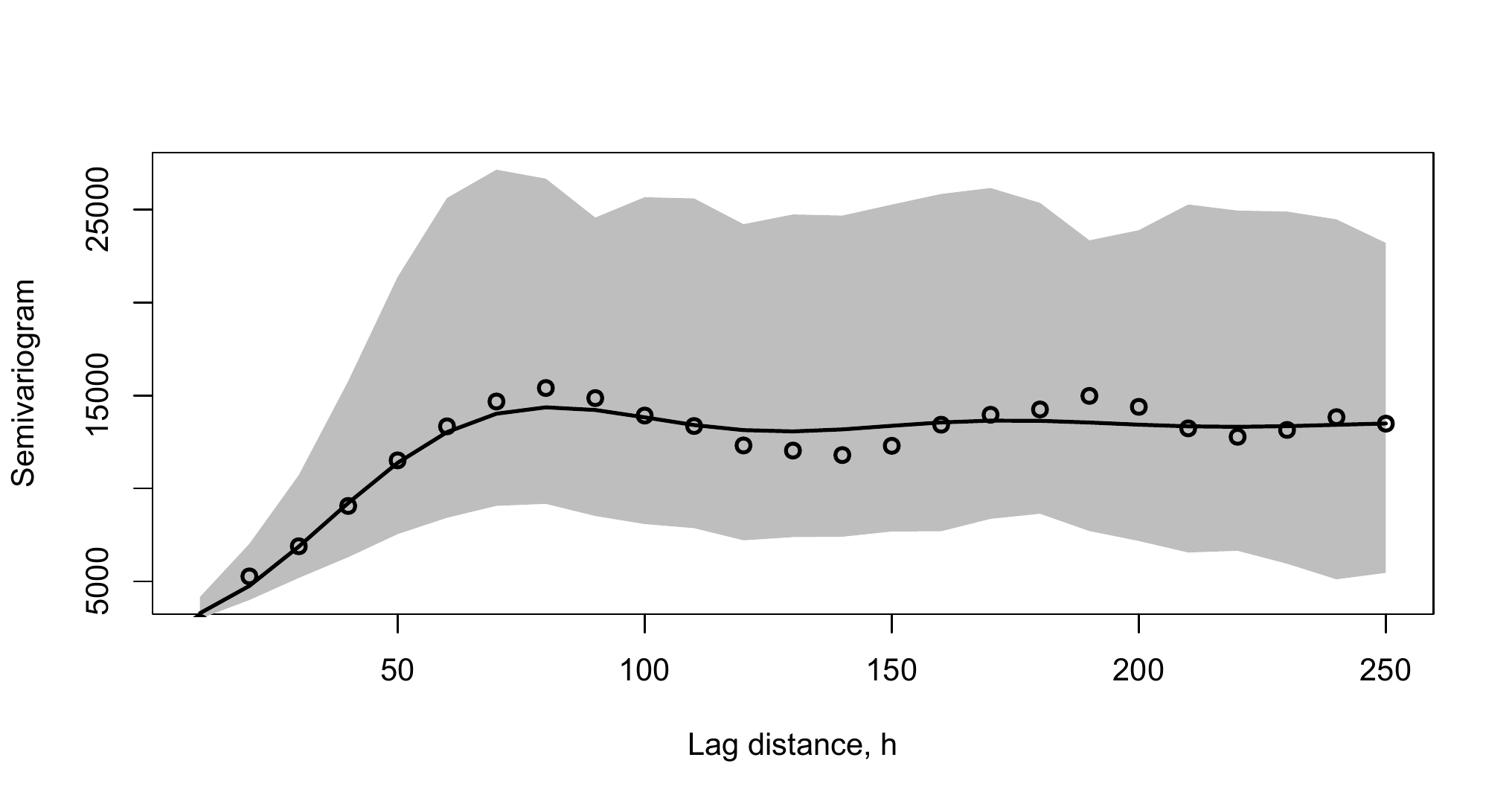}
\caption{
Comparison between the empirical semivariogram of the Swiss rainfall data (circles) and
the fitted sine-cosine wave model (solid line) together with 95\% simultaneous rank envelopes
(shaded areas) calculated from 39 simulations of the fitted model.}
\label{fig:4}
\end{center}
\end{figure}

Figure~\ref{fig:4} shows the empirical semivariogram for the data together with simulated pointwise 0.05 significance
envelopes obtained from 39 simulations of the sine-cosine wave model (such envelopes are obtained for each
value of $h$ by calculating the smallest and largest simulated values of $\hat{\gamma}(h)$; see Section 4.3.4 in \cite{moeller:waagepetersen:04}. We used the R package {\bf geoR} with some slightly modifications in its functions {\sf geoRCovModels}, {\sf cov.spatial} and {\sf grf} for the simulations.  For the sine-cosine wave model, $\hat{\gamma}(h)$
is within the shaded envelopes area for all value of $h$ which indicates that the sine-cosine wave  fits the data adequately.
\bibliographystyle{dcu}
\bibliography{Paperref}

\end{document}